# Optimally Tensile Strained $La_3Ni_2O_7$ Films as Candidate High-Temperature Superconductors on Designer $Ba_{1-x}Sr_xO$ (001) and SrO-$SrTiO_3$ Substrates

Liangliang Liu[1,2,#], Junhao Peng[3,#], Zhuangzhuang Qiao[1], Shuo Cai[1], Huafeng Dong[3,*], Yu Jia[1,2,*], and Zhenyu Zhang[4,*]

[1]*Key Laboratory for Special Functional Materials of Ministry of Education, School of Nanoscience and Materials Engineering, Henan University, Kaifeng 475004, China*

[2] *Institute of Quantum Materials and Physics, Henan Academy of Sciences, Zhengzhou 450046, China*

[3]*Guangdong Provincial Key Laboratory of Sensing Physics and System Integration Applications, School of Physics and Optoelectronic Engineering, Guangdong University of Technology, Guangzhou, Guangdong 510006, China*

[4] *International Center for Quantum Design of Functional Materials (ICQD), Hefei National Laboratory for Physical Sciences at Microscale (HFNL), University of Science and Technology of China, Hefei, Anhui 230026, China*

**Abstract**: Recent experiments have observed superconductivity up to 48 K in $La_3Ni_2O_7$-derived films under compressive strain imposed by the $SrLaAlO_4$ substrate, while such films on the $SrTiO_3$ substrate with tensile strain have failed to reach the superconducting state. Here we propose to broadly expand the choices of materials platforms to achieve high-$T_c$ superconducting $La_3Ni_2O_7$ films by proposing designer substrates of $Ba_{1-x}Sr_xO$ ($x$ = 0 - 1) that allow to continuously tune the strain in the films from being tensile to compressive. Our systematic study of the structural and electronic reconstructions of the strained $La_3Ni_2O_7$ bilayer film leads to the central finding that at the optimal tensile strain of ~2% ($x$ ~0.25), the spectral weight of the Ni $d_{z^2}$ orbital is peaked right at the Fermi level, and its hybridization with the Ni $d_{x^2-y^2}$ orbital is substantially enhanced. Consequently, the expected $T_c$ should be unprecedentedly high, at least substantially higher than those achieved in the compressive regime. Furthermore, our detailed thickness-dependent energetic analyses show that such films can be stably grown for thicknesses equal to or beyond the bilayer regime, and predict that the SrO-terminated $SrTiO_3$ should also be able to stabilize the films with optimal tensile strain and higher $T_c$'s.

The discovery of superconductivity with a critical temperature $T_c \approx 80$ K in the Ruddlesden-Popper (RP) bilayer phase of bulk nickelate La$_3$Ni$_2$O$_7$ under high pressure has sparked widespread interest in the field of high-$T_c$ superconductivity [1]. This discovery was soon followed by observations of superconductivity in the bilayer phase of La$_2$PrNi$_2$O$_7$ with $T_c = \sim 82.5$ K [2] and trilayer phase of La$_4$Ni$_3$O$_{10}$ with $T_c = \sim 30$ K [3]. Subsequent intensive efforts have been devoted to investigation of the electronic properties, density waves, and superconducting pairing [4-23]. However, the requirement of high pressure to maintain structural stability for bulk superconductivity poses extreme challenge for direct experimental probe of the superconducting phase. Recently, two independent research teams have achieved significant breakthroughs by synthesizing compressively strained La$_3$Ni$_2$O$_7$-derived films on the SrLaAlO$_4$ substrate, with $T_c$ exceeding 40 K [24-27]. These latest developments demonstrate that substrate strain can provide unprecedented opportunities to realize ambient-pressure high-$T_c$ superconductivity in nickelate superconductors, as well as to explore the electronic structures and pairing mechanisms.

Despite the distinct advances, various controversies remain about the systems. In particular, angle-resolved photoemission spectroscopy (ARPES) measurements on the strained La$_{2.85}$Pr$_{0.15}$Ni$_2$O$_7$ films reported a diffuse γ band originating from the Ni bonding $3d_{z^2}$ orbital, resembling the pressure-induced effect in bulk systems [28,29]. On the other hand, ARPES investigations of the strained La$_2$PrNi$_2$O$_7$ films showed the bonding $3d_{z^2}$ orbital located ~70 meV below the Fermi level ($E_F$) [30]. Additionally, theoretical studies suggested that both compressive and tensile strain can effectively tune the electronic structure of the La$_3$Ni$_2$O$_7$ films that favor superconductivity [31,32]. However, all heteroepitaxial systems that have exhibited high-$T_c$ superconductivity so far are confined to the compressive strain regime imposed by only the SrLaAlO$_4$ substrate [24-27], limiting the ability to manipulate the electronic structure and directly probe the connection between the γ bands and superconductivity. Subsequent attempts using the tensile strain imposed by the SrTiO$_3$ substrate have failed to induce superconductivity [33,34]. It is thus crucial to identify practical substrates that allow to continuously tune the strain from being compressive to tensile, and to clarify the role of the γ bands in superconductivity as well as to reach higher $T_c$'s.

In this Letter, we propose to broadly expand the choices of materials platforms to achieve high-$T_c$ superconducting La$_3$Ni$_2$O$_7$ films by proposing designer substrates of

Ba$_{1-x}$Sr$_x$O ($x$ = 0 - 1) that allow to continuously tune the strain in the films from being tensile to compressive. Using first-principles calculations with the ubiquitous van der Waals forces included, we systematically explore the role of tunable epitaxial strain in driving the structural and Ni $d_{z^2}$ orbital evolutions of the films. The central finding is that at the optimal tensile strain of ~2% ($x \sim 0.25$), the spectral weight of the Ni $d_{z^2}$ orbital is peaked right at the Fermi level ($E_F$), and its hybridization with the Ni $d_{x^2-y^2}$ orbital is substantially enhanced. Consequently, the expected $T_c$ should be unprecedentedly high, at least substantially higher than those achieved in the compressive regime. Furthermore, our detailed thickness-dependent energetic analyses show that such films can be stably grown for thicknesses equal to or beyond the bilayer regime, and predict that the SrO-terminated SrTiO$_3$ should also be able to stabilize the films with optimal tensile strain and higher $T_c$'s. This versatile tunability in the substrate-imposed strain will offer critical insights into the key factors influencing superconducting pairing in Ni-based superconductors and expand the high-temperature superconducting landscape of nickelate thin films.

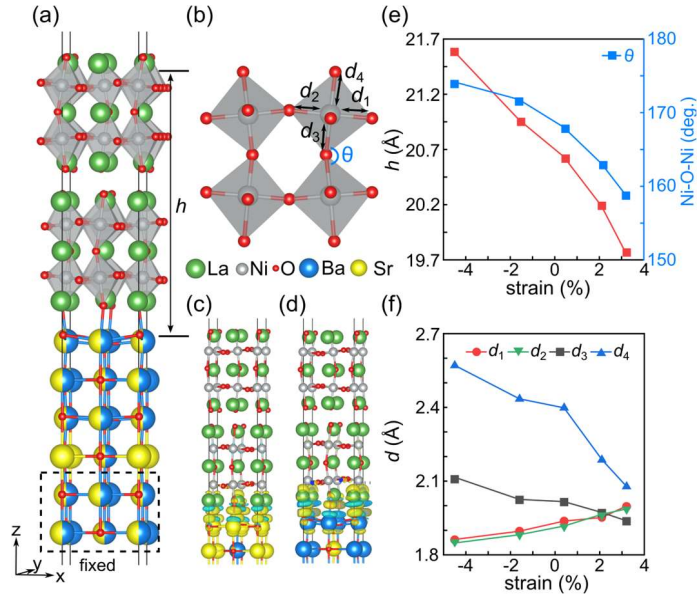

FIG. 1. (a) Constructed La$_3$Ni$_2$O$_7$/Ba$_{1-x}$Sr$_x$O (001) heterostructures. (b) Sketch of a NiO$_6$ octahedron, with $d_1$, $d_2$, $d_3$, and $d_4$ labeling the corresponding Ni-O bond lengths, and $\theta$ represents the Ni-O-Ni bond angle along the $c$ axis. Charge density differences $\Delta\rho$ ($\Delta\rho$ = $\rho_{system} - \rho_{La_3Ni_2O_7} - \rho_{substrate}$) of heterostructure (c) La$_3$Ni$_2$O$_7$/Ba$_{0.25}$Sr$_{0.75}$O and (d) La$_3$Ni$_2$O$_7$/Ba$_{0.75}$Sr$_{0.25}$O. Isosurface ±0.015 e/Å$^3$: the yellow (blue) color denotes electron accumulation (depletion). (e) Strain-dependent height $h$ along the $c$ axis, Ni-O-Ni bond angle $\theta$, and (f) the Ni-O bond lengths of the strained La$_3$Ni$_2$O$_7$ films.

At ambient pressure, bilayer La$_3$Ni$_2$O$_7$ adapts the structure of the orthorhombic *Amam* space group, characterized by Ni-O-Ni bond angle distortion along *c* axis ($\theta \approx 168.0°$) [1]. The optimized *Amam* structure yields lattice parameters [bulk structure in Fig. S1(a) [35]] that closely match the experimental values [10]. The electronic states around the $E_F$ mainly come from the contribution of the Ni $d_{x^2-y^2}$ and $d_{z^2}$ orbitals, as shown in Fig. S1(b) [35], and interlayer coupling via the apical O sites splits the Ni $d_{z^2}$ orbital into bonding ($\gamma$) and antibonding ($\gamma_{an}$) bands with a gap of ~ 0.2 eV. The mixed $d_{x^2-y^2}$ and $d_{z^2}$ orbitals form two types of electron-pocket bands ($\alpha$ and $\beta$), crossing the $E_F$ [Fig. S1(c) [35]]. As reference systems, we first study the evolution of structural and electronic properties of bulk La$_3$Ni$_2$O$_7$ under compressive or tensile strains, and find that pronounced tensile strain can metallize the $\gamma$ bands with large electronic occupation at $E_F$ [Figs. S2-S4 [35]], as observed previously [32]. In order to effectively impose strain including compressive and tensile effects for the bilayer La$_3$Ni$_2$O$_7$ experimentally, we design suitable candidate materials Ba$_{1-x}$Sr$_x$O ($x$ = 1, 0.75, 0.5, 0.25 and 0) as substrates to construct heterostructures of bilayer La$_3$Ni$_2$O$_7$ films [Fig. 1(a)]. Here Ba$_{1-x}$Sr$_x$O adapts the cubic structures, coming from the original cubic SrO and BaO ($Fm\bar{3}m$ space group) [40, 41] by tuning the Sr and Ba ratio [bulk and surface configurations in Fig. S5 [35]. The in-plane lattice constants range from 5.167 Å (SrO) to 5.582 Å (BaO), and the corresponding strain ranges from ~ -4.5% to 3.2% onto the La$_3$Ni$_2$O$_7$ films (pseudo-tetragonal $a_p = (a^2 + b^2)^{1/2}/\sqrt{2} = 5.409$ Å). Specifically, substrates SrO, Ba$_{0.25}$Sr$_{0.75}$O, Ba$_{0.5}$Sr$_{0.5}$O, Ba$_{0.75}$Sr$_{0.25}$O, and BaO (001) produce ~ -4.5%, -1.6%, 0.4%, 2.1%, and 3.2% strain for La$_3$Ni$_2$O$_7$ films, respectively.

Fig. 1(a) shows the structural model of a bilayer La$_3$Ni$_2$O$_7$ film on Ba$_{1-x}$Sr$_x$O substrates, and the film height $h$ refers to the distance between the top surface layer of the film and the substrate. Figs. 1(e) and 1(f) show the evolution of the lattice structure, tilting angle $\theta$, and Ni-O bond lengths [marked in Fig. 1(b)] of the strained La$_3$Ni$_2$O$_7$ films on different substrates. It is noted that compressive strain (e.g., Ba$_{0.25}$Sr$_{0.75}$O: -1.6%, SrO: -4.5%) suppresses the octahedral tilting ($\theta \approx 172°, 174°$), and expands the film height $h$ (20.96 Å, 21.58 Å), which significantly deviate from $c$ = 19.73 Å of bulk phase at 30 GPa (Fig. S6 [35]). Similar conclusion can be drawn concerning the bond lengths $d_3$ and $d_4$ between the Ni and apical oxygens, which increase to 2.023 Å and 2.433 Å at -1.6% respectively, much larger than the in-plane bond lengths $d_1$ = 1.896 Å and $d_2$ = 1.881 Å [Fig. 1(f)], indicating that compressive strain drives the *c*-axis to further elongate. Accordingly, the charge hybridization of the Ni $d_{z^2}$ orbital with the apical O $p_z$ orbital is weakened, as shown in the charge density distribution shown in Fig. S7 [35]. In contrast, the $Ba_{0.5}Sr_{0.5}O$, $Ba_{0.75}Sr_{0.25}O$ and BaO substrates all impose tensile strains, which drives the $h$ value to continuously decrease to 20.63, 20.18 and 19.75 Å, respectively. In particular, the $h$ values on the $Ba_{0.75}Sr_{0.25}O$ and BaO substrates tend to be close to the lattice constant $c$ = 19.734 Å of the bulk sample obtained at 30 GPa. At the same time, $d_3$ and $d_4$ decrease on these substrates, and on the $Ba_{0.75}Sr_{0.25}O$ (BaO) substrate, the $c$-axis of the $NiO_6$ octahedrons is suppressed significantly as the $d_3$ and $d_4$ reduced to 1.972 Å (1.938 Å) and 2.187 Å (2.078 Å) respectively. Again, these variations are similar to those ($d_3$ = 1.901 Å and $d_4$ = 2.126 Å) of the bulk sample under 30 GPa. Tensile strain in turn also enhances charge hybridization between the Ni $d_{z^2}$ and O $p_z$ orbitals [Fig. S7 [35]]. These calculated results demonstrate that strain effect could profoundly modify the lattice constants, the $NiO_6$ octahedral shape, and charge distribution, thereby renormalizing the electronic structure of bilayer $La_3Ni_2O_7$.

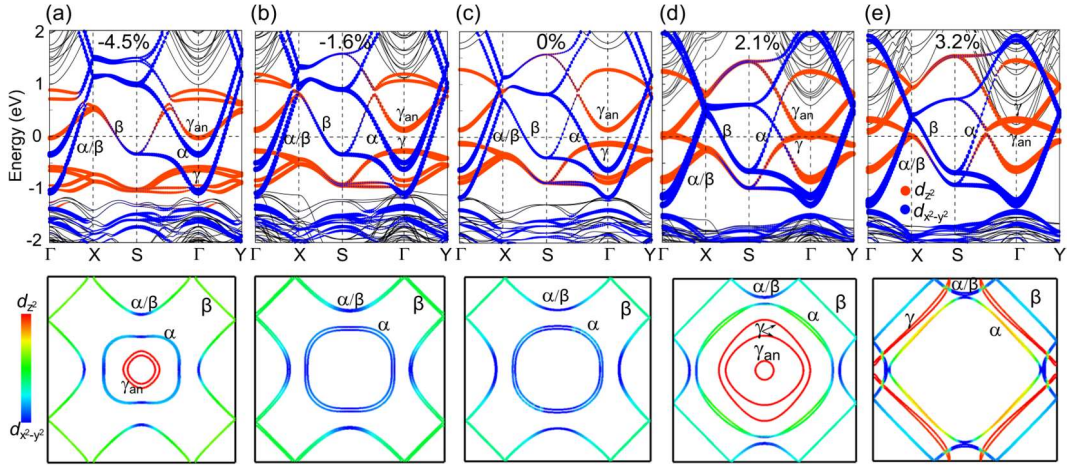

FIG. 2. Calculated band structures of bilayer $La_3Ni_2O_7$ films on substrates: (a) SrO (-4.5% strain), (b) $Ba_{0.25}Sr_{0.75}O$ (-1.6% strain), (d) $Ba_{0.75}Sr_{0.25}O$ (2.1% strain) and (e) BaO (3.2% strain), as compared to (c) those of free-standing bilayer $La_3Ni_2O_7$ film (without strain). Orbital projection: Ni $d_{z^2}$ (red), $d_{x^2-y^2}$ (blue). Corresponding Fermi surfaces colored by Ni ($d_{z^2}$ − $d_{x^2-y^2}$)/($d_{z^2}$ + $d_{x^2-y^2}$).

We next examine the evolution of charge density, band structure, and the Fermi surface of the strained bilayer $La_3Ni_2O_7$ films on the proposed substrates. We note because $Ba_{1-x}Sr_xO$ are non-polar substrates, irrespective of compressive or tensile effects given by $Ba_{0.25}Sr_{0.75}O$ and $Ba_{0.75}Sr_{0.25}O$, the charge transfer is mainly surrounding interfacial La-O layer of film and Sr-O/Ba-O of substrates, and its value is very small, which means that no significant interfacial charge transfer is observed between the substrates and films [Figs. 1(c) and 1(d)]. Thus, the substrates do not influence the Ni valence states of the films, similar to the case of SrLaAlO$_4$ (001) that do not significantly induce charge transfer into the La$_3$Ni$_2$O$_7$ thin film [32], maintaining its electronic structure and potentially superconducting properties of the overlayer.

Band structures in Fig. 2 reveal distinct strain-dependent responses. Compared to the unstrained film [Fig. 2(c)], due to compressive strains (Ba$_{0.25}$Sr$_{0.75}$O or SrO substrate), the *c*-axis of the NiO$_6$ octahedrons is elongated, while the Ni $d_{z^2}$-derived $\gamma$ bands tend to move towards the lower energy level and resist metallization [Figs. 2(a), 2(b)]. For the SrO substrate, the pronounced compression even lowers the antibonding $\gamma_{an}$ bands across the $E_F$ with an unconventional occupation around the $\Gamma$ point, forming two Fermi sheets [Fig. 2(a)]. Conversely, tensile strain leads to shortening of the *c*-axis and enhanced tilting of the NiO$_6$ octahedrons, while Ni $d_{z^2}$ orbitals are lifted towards the $E_F$. Especially, on the Ba$_{0.75}$Sr$_{0.25}$O or BaO substrate, larger tensile strain drives a net electron transfer from the Ni $d_{z^2}$ to the Ni $d_{x^2-y^2}$ orbital with hole doping into the $\gamma$ bands that become metallized, as shown in Figs. 2(d) and 2(e). Additionally, due to the substrate effect, the $\gamma_{an}$ bands are also lowered across $E_F$, which is different from the bulk case under 2% strain that only the $\gamma$ bands cross the $E_F$ [Fig. S2 [35]]. The BaO substrate imposes the largest tensile strain of 3.2%, and the $\gamma$ bands evolve across the Brillouin zone boundary, similar to the bulk case under 3% tensile strain [Fig. S2 [35]]; therefore, the $\gamma$ bands no longer form the closed Fermi sheets, exhibiting complex Fermi surface topology [Fig. 2(e)]. Crucially, the $\gamma$ bands achieve effective metallization with the highest electronic occupation at $E_F$ under the optimal tensile strain imposed Ba$_{0.75}$Sr$_{0.25}$O.

Previous research reported that electronic states of the Ni $d_{z^2}$ and Ni $d_{x^2-y^2}$ orbitals at $E_F$ dominate the superconducting pairing [1-3], and strain-induced distortions play a crucial role by reshaping the octahedral crystal field splitting and orbital occupation. As shown in Figs. 3(a)-3(e), modulating the strain sign and strength shifts the orbital competition: under strong compressive strain, $d_{z^2}$ and $d_{x^2-y^2}$ contributes to $N_{E_F}$ are nearly equally [Fig. 3(f)]. As strain transits from compressive to tensile, the $d_{z^2}$ contribution increases progressively. In particular, under tensile strain, $d_{z^2}$ becomes dominant, significantly surpassing $d_{x^2-y^2}$. Most significantly, at 2.1% tensile strain [Figs. 3(d) and 3(f)], both the $d_{z^2}$ orbital and total DOS peak near the $E_F$, a behavior highly consistent with the recently reported superconducting property in high-pressure La$_3$Ni$_2$O$_7$ [1,42]. This $d_{z^2}$-orbital-dominated electronic structure favors interlayer pairing, creating favorable conditions for achieving higher critical temperatures. Consequently, experimental strategies utilizing substrates to induce ~2.1% tensile strain are expected to significantly enhance the $T_c$ in the Ni-based thin film samples.

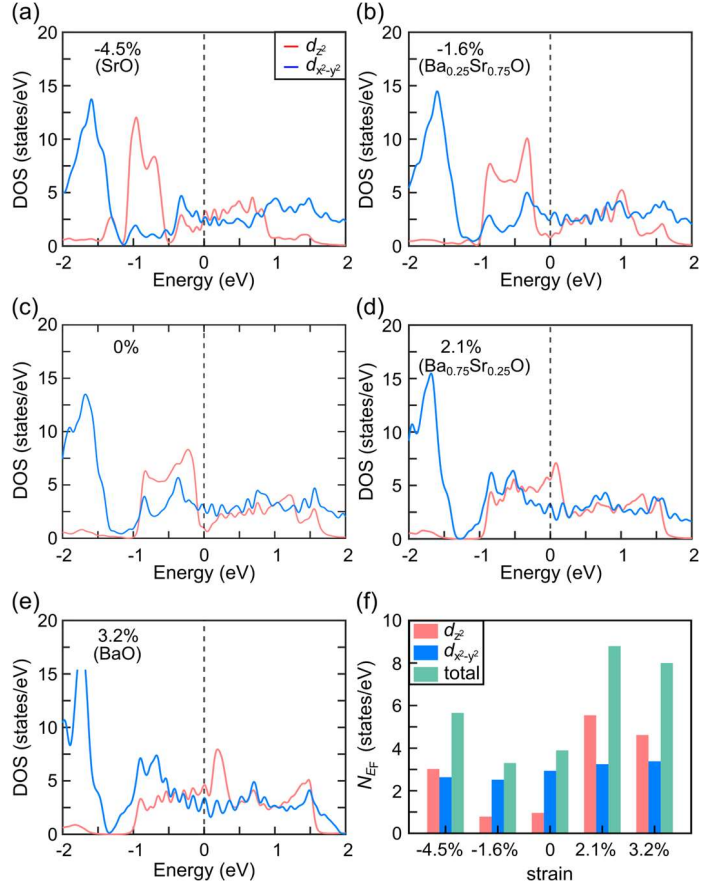

Fig. 3. (a)-(e) Calculated density of states (DOS) of a La$_3$Ni$_2$O$_7$ bilayer on different substrates. (f) DOS at $E_F$ ($N_{E_F}$) for the Ni $d_{z^2}$ and $d_{x^2-y^2}$ orbitals.

Next, we examine the layer-by-layer growth of La$_3$Ni$_2$O$_7$ films on the designed substrates, and compare their stability on several substrates currently attempted in experiments, including SrLaAlO$_4$, SrTiO$_3$ with TiO$_2$ termination (TiO$_2$-SrTiO$_3$) and SrO termination (SrO-SrTiO$_3$) [Fig. 4(a)]. The stability of a thin film with a given thickness on the substrate can be judged from the second derivative $\Delta^2\overline{E}_N$ of average total energy $\overline{E}_N$ [45,46]. Figs. 4(b) and 4(c) present the calculated $\overline{E}_N$ and $\Delta^2\overline{E}_N$ for La$_3$Ni$_2$O$_7$ films up to 6 ML on various substrates. For 1ML epitaxial growth, the La$_3$Ni$_2$O$_7$ film exhibits high stability on Ba$_{0.25}$Sr$_{0.75}$O, Ba$_{0.75}$Sr$_{0.25}$O, SrLaAlO$_4$ and SrO-SrTiO$_3$ substrates, behaving like a wetting layer [Fig. 4(c)]. Stable bilayer (2 ML) growth is also achievable on these four substrates. Notably, the bilayer La$_3$Ni$_2$O$_7$ film on the predicted Ba$_{0.75}$Sr$_{0.25}$O substrate shows particularly enhanced stability. Theoretically, stable growth up to 6 ML remains possible on these four substrates. However, stable La$_3$Ni$_2$O$_7$ bilayers cannot be obtained on TiO$_2$-SrTiO$_3$ due to the distortion of TiO$_6$ octahedral units of the substrate at the interface [Fig. 4(d)], putting aside the fact that the oxygen vacancies are easily formed on TiO$_2$ terminated surfaces in practical situations. Notably, for our designed substrate Ba$_{0.75}$Sr$_{0.25}$O, the interfacial structure does not exhibit such distortion under similar tensile condition of ~2% [Fig.4(e)]. Overall, we have identified superior substrates Ba$_{0.75}$Sr$_{0.25}$O and SrO-SrTiO$_3$ for growing stable La$_3$Ni$_2$O$_7$ superconducting films, which warrants experimental verification.

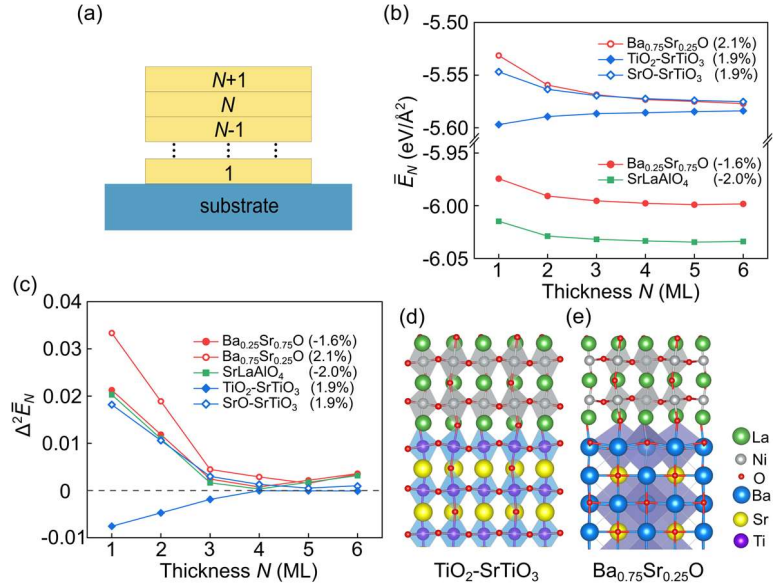

Fig. 4: (a) Sketch of continuous La$_3$Ni$_2$O$_7$ thin films on a substrate. Here a monolayer (ML) means half of a bilayered unit in bulk La$_3$Ni$_2$O$_7$, and $N$ represents the number of monolayers. (b) Average total energy ($\bar{E}_N$) per unit area of the La$_3$Ni$_2$O$_7$ films on different substrates and (c) its second derivative $\Delta^2\bar{E}_N$ [45,46]. Optimized interlayer configurations of (d) La$_3$Ni$_2$O$_7$/TiO$_2$-SrTiO$_3$ and (e) La$_3$Ni$_2$O$_7$/Ba$_{0.75}$Sr$_{0.25}$O.

On the physical feasibility side, we note that experiments have already synthesized BaO, SrO substrates [47-49], and their alloyed forms such as Ba$_{0.7}$Sr$_{0.3}$O [43,44], as well as SrO-SrTiO$_3$ substrate [50,51]. In particular, such Ba$_{0.7}$Sr$_{0.3}$O and SrO-SrTiO$_3$ substrates can impose the nearly optimal tensile strain of approximately 2% to the La$_3$Ni$_2$O$_7$ film. Additionally, for the designer substrate Ba$_{1-x}$Sr$_x$O, based on these experimental studies, continuously tuning the Ba and Sr relative ratio would provide an effective way to modulate the in-plane lattice constant of the substrate, thereby enabling continuous control of strain in the film from being compressive to tensile. As previous studies have demonstrated [52], a high-throughput approach for synthesizing substrates can be used to prepare a series of $Ba_{1-x}Sr_xO$ substrates, on which the $La_3Ni_2O_7$ films can be fabricated.

To date, superconductivity in $La_3Ni_2O_7$ films has only been observed under compressive strain of $SrLaAlO_4$, and the strain imposed by the substrates $Ba_{1-x}Sr_xO$ could effectively regulate the distortion of the $NiO_6$ octahedron in the film samples and the relative contribution of the $d_{z^2}$ and $d_{x^2-y^2}$ orbitals near the $E_F$, as well as drive the transition of the $d_{z^2}$ bonding states from being non-metallic to metallic. Experimentally, by comparing the superconducting and electronic properties enabled by the versatile strain tunability of the $Ba_{1-x}Sr_xO$ substrates, critical insights can be drawn into the key factors of the $d_{z^2}$ and $d_{x^2-y^2}$ orbitals in influencing superconducting pairing of Ni-based superconductors. Furthermore, at the optimal tensile strain of ~2% ($x$ = ~0.25), the $d_{z^2}$ and $d_{x^2-y^2}$ orbitals could contribute maximally to the electronic occupancy at the $E_F$, driving the $La_3Ni_2O_7$ film as a higher-$T_c$ candidate superconductor than those achieved in the compressive regime.

In summary, we have designed a series of $Ba_{1-x}Sr_xO$ substrates across the alloying concentration $x$, and systematically explored the role of tunable epitaxial strain in driving the structural and electronic reconstructions. Our systematic study of the strained $La_3Ni_2O_7$ bilayer film leads to the central finding that at the optimal tensile strain of ~2% ($x$ ~0.25), the spectral weight of the Ni $d_{z^2}$ orbital is peaked right at the Fermi level, and its hybridization with the Ni $d_{x^2-y^2}$ orbital is substantially enhanced. Consequently, the expected $T_c$ should be unprecedentedly high, at least substantially higher than those achieved in the compressive regime. Furthermore, our detailed thickness-dependent energetic analyses show that such films can be stably grown for thicknesses equal to or beyond the bilayer regime, and predict that the $SrO-SrTiO_3$ is also able to stabilize the films with optimal tensile strain and higher $T_c$'s. Collectively, these central findings provide critical insights into the delicate roles of structural and electronic reconstructions in superconducting pairing.

We acknowledge the support from the Innovation Program for Quantum Science and Technology (Grant No. 2021ZD0302800), the National Natural Science Foundation of

China (Grants No. 12104129 and No. 12074099) and Natural Science Foundation of Henan Province (Grants No. 242300421162 and 242300421213).

[#]These authors contributed equally to this work

*Corresponding authors: Huafeng Dong (hfdong@gdut.edu.cn); Yu Jia (jiayu@henu.edu.cn); Zhenyu Zhang (zhangzy@ustc.edu.cn)